\documentclass[12pt,tightenlines,nofootinbib,aip]{revtex4}
\usepackage{amssymb}
\usepackage{amsmath}
\usepackage{amsfonts}
\usepackage[stable]{footmisc}
\usepackage{hyperref}

\begin{document}

\title{Rethinking the scientific enterprise: in defense of reductionism}

\author{Ian T. Durham}
\email[]{idurham@anselm.edu}
\affiliation{Department of Physics, Saint Anselm College, Manchester, NH 03102}
\date{\today}

\begin{abstract}
In this essay, I argue that modern science is \emph{not} the dichotomous pairing of theory and experiment that it is typically presented as, and I offer an alternative paradigm defined by its functions as a human endeavor.  I also demonstrate how certain scientific debates, such as the debate over the nature of the quantum state, can be partially resolved by this new paradigm.
\end{abstract}

\maketitle

\begin{quote}I have begun to enter into companionship with some few men who bend their minds to the more solid studies, rather than to others, and are disgusted with Scholastic Theology and Nominalist Philosophy.  They are followers of nature itself, and of truth, and moreover they judge that the world has not grown so old, nor our age so feeble, that nothing memorable can again be brought forth. --Henry Oldenberg, as quoted in~\cite{Boorstin:1983vn}.
\end{quote}

\section{}
Science is a living, breathing -- and very human -- enterprise.  As such, it has always been a malleable process.  Indeed, that is one of its enduring traits: not only does science prescribe a system by which its predictions may be refined by additional knowledge, but its \emph{very nature} changes as our understanding of the world and ourselves broadens.  Nevertheless, there is an  over-arching paradigm to modern science whose origins are rooted in the works of Alhazen\footnote{Ab\={u} `Al\={i} al-\d{H}asan ibn al-\d{H}asan ibn al-Haytham (965 CE -- c. 1040 CE), also known as Ibn al-Haytham and sometimes al-Basri.} who flourished during the Islamic Golden Age, circa 1000 CE.  In its simplest form, this paradigm consists of the posing of questions and the subsequent testing of those questions~\cite{Sambursky:1974fk}.  This process is, of course, cyclic as the testing of the original questions very often leads to new ones.  But the asking of a question is really at the root of all scientific endeavor and stems from humanity's innate curiosity about itself and the world around us.  In a sense, we all remain children, continually asking `why?'  In more modern scientific terms, the act of questioning forms the basis of a scientific theory that is ``a well-substantiated explanation of some aspect of the natural world, based on a body of facts that have been repeatedly confirmed through observation and experiment''~\cite{Science:1999bh}.  In other words, Alhazen's paradigm breaks science into two equal parts: theory and experiment.

While there have been modern refinements to Alhazen's basic framework, notably the adoption of the hypothetico-deductive\footnote{The term `hypothetico-deductive' has been attributed to William Whewell, though evidence for this is lacking as the term does not appear in any of his works on the inductive sciences.} model, the basic division into theory and experiment remains.  Victoria Stodden has recently proposed that computational science be recognized as a third division and, indeed, this is an attractive suggestion~\cite{Stodden:2010kx}.  But it would fail to address certain persistent problems with both theory and experiment that raise deeper questions about the overall methodology of science.  Clues to a solution to these problems can be found in the origins of that methodology.

While a precise formulation of the history of modern scientific methodology is not only lengthy but somewhat subjective, it is generally agreed that the revolution it sparked began in 17th century Europe and many of its principles were codified in the early documents and practices of the Royal Society of London, arguably the worldÕs oldest scientific organization\footnote{The history of the Royal Society is tightly linked with a number of organizations that arose in the mid--17th century including Acad\'{e}mie Monmor, the Acad\'{e}mie des sciences, and Gresham College~\cite{Boorstin:1983vn}}~\cite{Boorstin:1983vn}.  As Thomas Sprat wrote, the Royal Society's purpose was ``not the Artifice of Words, but a bare knowledge of things'' expressed through ``Mathematical plainness''~\cite{Boorstin:1983vn}.  This early scientific community developed a highly mechanistic approach to science that, while applied with equal vigor to anything tangible (and thus encompassing the modern fields of astronomy, chemistry, biology, physiology, et. al.), was decidedly grounded in the \emph{physical}.  The modern field that we recognize as physics has been called ``the most fundamental and all-inclusive of the sciences''~\cite{Feynman:1963ys}.  Arguably a portion of that inclusivity stems from the fact that all the other sciences are constrained by physical laws.  This is one way in which scientific reductionism can be interpreted --- a `reduction' of the other sciences to physics.  But physics is also inclusive by dint of its methods.  Physics, throughout its history, has hewed most closely to the mechanistic approach developed in the 17th century and, indeed, this is the other way in which scientific reductionism is traditionally interpreted --- a `reduction' of a \emph{system} to its constituent parts in an effort to better comprehend the whole.  

This interpretation of reductionism is closely related to the notion of causality and, as a view of science, has been challenged in recent years as a result of work on emergence and complex systems~\cite{Ulanowicz:1997zr,Andersen:1972ly,Kauffman:2006ve,Lehrer:2011qf}.  As Jonah Lehrer\footnote{The ideas for the present essay were in large part developed as a rejoinder to Lehrer \emph{prior} to his resignation from the \emph{New Yorker} after admitting to fabricating quotes.  That incident should have no bearing on what is written and discussed here.} wrote in a recent article
\begin{quote}
[t]his assumption --- that understanding a systemÕs constituent parts means we also understand the causes within the system --- is not limited to the pharmaceutical industry or even to biology. It defines modern science. In general, we believe that the so-called problem of causation can be cured by more information, by our ceaseless accumulation of facts. Scientists refer to this process as reductionism. By breaking down a process, we can see how everything fits together; the complex mystery is distilled into a list of ingredients~\cite{Lehrer:2011qf}.
\end{quote}
Lehrer's article, however, focused almost exclusively on a single aspect of scientific methodology that is not necessarily mechanistic and that is misunderstood, even by scientists themselves: statistics and mathematical modeling.  If reductionism is indeed what Lehrer claims it is, then statistical methods and mathematical modeling are most definitely \emph{not} reductionist since they only seek to find mathematical structures that explicitly match existing data.  This point is perhaps the most misunderstood in all of science.  As an example, we consider first the relationship between statistics and probability.

\section{}
Statistics often accompanies probability (at least in textbook titles and encyclopedia entries).  But this belies a subtle but important difference between the two.  Both are indeed disciplines in their own right that fall under the larger umbrella of mathematics and logic.  But only statistics is an actual \emph{tool} of science.  Probability is a logico-mathematical description of random processes.  Statistics, on the other hand, is a methodology by which aggregate or `bulk' information may be analyzed and understood.  It loses its meaning and power when applied to small sample sizes.  And there's the rub.  If reductionism is the act of breaking down a process in order to understand its constituent parts, as Lehrer claims, statistics is the \emph{antithesis} of reductionism because it makes no such effort.

Why then do we stubbornly persist in thinking that statistical methods in science can masquerade as some kind of stand-in for reductionism?  Why do we expect more from statistics than we have a right to?  Statistics is a very --- \emph{very} --- important tool in science, but it is often misapplied and its results are often misinterpreted.  Few understood this better than E.T. Jaynes.  Jaynes spent the better part of his career attempting to correct one of the more egregious misconceptions, one that is intimately related to the difference between probability and statistics.  

Roughly speaking, statistics generally describe information we \emph{already know} or data we've \emph{already collected}, whereas probability is generally used to \emph{predict what might happen in the future}.  As Jaynes astutely noted, if we imagine data sampling as an exchangeable sequence of trials,
\begin{quote}
the probability of an event at one trial is not the same as its frequency in many trials; but it is numerically equal to the\emph{expectation} of that frequency; and this connection holds whatever correlation may exist between different trials \ldots The probability is therefore the ``best'' estimate of the frequency, in the sense that it minimizes the expected square of the error~\cite{Jaynes:1989fk}.
\end{quote}
In other words, probabilities can only be \emph{accurately} formulated from statistical data if that data arose from a \emph{perfectly repeatable} series of experiments or observations.  This is the genesis of the interpretational debate over the meaning of the word `probability,' with the frequentists on one side claiming a probability assignment is really nothing more than an assignment of the frequency of occurrence of a given outcome of a trial, and the Bayesians on the other side claiming a probability assignment is a state of knowledge.  As Jaynes clearly notes, the frequency interpretation is only valid under strictly enforceable conditions whereas the Bayesian view is more general.

What does the Bayesian interpretation of probability tell us about reductionism?  The key to the Bayesian interpretation is the notion that, if probabilities represent our states of knowledge, measurements \emph{update} these states of knowledge.  Thus knowledge is gained in an incremental manner\footnote{This is not necessarily the same thing as \emph{sequential}, as is clearly demonstrated by certain quantum states.} which is the essence of reductionism.  Thus probabilities, in a Bayesian context, are absolutely reductionist.  As Jaynes points out, it \emph{is} possible to give probabilities a frequentist interpretation, in which case they connect to the more aggregate descriptions provided by statistics, but only under certain strict conditions.

All of this does not necessarily obviate the need for the broader generalizations provided by statistics.  In fact, as the foundational basis for thermodynamics, statistics as understood in the sense of distributions of measured quantities, has been very successful in explaining large-scale phenomena in terms of the bulk behavior of microscopic processes.  Similar arguments can be made in terms of fluid dynamics, atmospheric physics, and similar fields.  As Jaynes pointed out,
\begin{quote}
[i]n physics, we learn quickly that the world is too complicated for us to analyze it all at once.  We can make progress only if we dissect it into little pieces and study them separately.  \emph{Sometimes, we can invent a mathematical model which reproduces several features of one of these pieces, and whenever this happens we feel that progress has been made}~\cite{Jaynes:1998uq}, [emphasis added].
\end{quote}
Thus statistics is one of the primary methods by which larger-scale patterns are discovered.  These patterns thus \emph{emerge} in aggregate behavior from the underlying pieces.  However, it is wrong to assume that such patterns can emerge \emph{completely independently} of the underlying processes.  This is tantamount to assuming that macroscopic objects can exist independently of their underlying microscopic structure.  The melting of an ice cube clearly refutes this notion.  

Of course, very few true anti-reductionists would argue this fairly extreme view.  Instead they argue an intermediate position such as that proposed by P.W. Andersen~\cite{Andersen:1972ly}.  Andersen fully accepts reductionism, but argues that new principles appear at each level of complexity that are not merely an extension of the principles at the next lower level of complexity.  In another words, Andersen is suggesting that were we to be endowed with a sufficiently powerful computer \emph{and} were we to have a full and complete understanding of, say, particle physics, we \emph{still} would not be able to `derive' a human being, for example, or, at the very least, the basic biological laws governing human beings.  Biology and chemistry, to Andersen, are more than just applied or extended physics.  This is precisely the point Lehrer is trying to make.  But there are two fundamental problems with this argument.  

The first problem is that this assumes that no amount of additional knowledge can bridge the gap between levels of complexity, i.e. it takes as \emph{a priori} that reductionism (or `constructionism,' as Andersen calls it) is either wrong or incomplete.  But this is \emph{logically unprovable}.  As Carl Sagan wrote, ``[y]our inability to invalidate my hypothesis is not at all the same thing as proving it true''~\cite{Sagan:1996uq}.  In fact, this is precisely the same argument that proponents of creationism and intelligent design employ in claiming the universe (especially biological life) is too complex to arise from simpler, less complex rules~\cite{Atkins:2006fk}.  

To understand the second problem with the anti-constructionist view, as I will call it, consider two physical systems, $X$ and $Y$, each independently described by the same set of mathematical structures, $M$, that we take to be the minimum set that fully describes each system\footnote{We are inherently assuming, here, that mathematics \emph{can} fully describe physical systems.  This may or may not be true, but for now we assume that it is.}.  Now suppose that completely combining these physical systems gives rise to a \emph{third} physical system, $Z$, that is described by a set of mathematical structures, $N$, where $M\ne N$ and $N$, like $M$, is taken to be the minimum set of structures that fully describes the system (in this case, $Z$).  In this scenario, $X$ and $Y$ are more `fundamental' than $Z$ and thus $M$ must necessarily be a more restrictive set of structures than $N$.  The anti-constructionist view assumes that $Z$ is more complex than merely the combination of $X$ and $Y$. This then implies that $N$ cannot be derived from $M$ alone.  In fact, it implies that there are structural elements of $N$ that cannot be derived from \emph{any} more primitive set of structures.  But mathematics is built on logic and is thus internally completely self-consistent.  In other words, mathematics is and always has been \emph{assumed} to be purely reductionist.  Thus, if anti-constructionism is correct then this assumption about mathematics is wrong. But no evidence of discord within mathematics exists.  So why, then, is there so much discord of this nature within science?

\section{}
Recall that Alhazen's paradigm breaks science into two equal parts: theory and experiment.  In this paradigm, experiments `describe' the universe and theories `explain' it.  In this light, consider the development of Newtonian gravity in the 17th century.  We can assign Galileo the role of experimenter/observer for his work with falling bodies, bodies on an inclined plane, and his observations of the moons of Jupiter, the latter of which importantly showed that celestial objects could orbit other celestial objects aside from the earth.  This final point emphasizes the fact that a full theory of gravity had to take into account the movement of \emph{celestial} bodies as well as terrestrial ones.  Where, then, in this historical context, can we place Kepler?  The data used by Kepler in the derivation of his three laws of planetary motion was largely taken by Tycho Brahe.  They were not \emph{explained} until nearly six decades after Kepler's death in 1630 when Newton published his \emph{Philosophi{\ae} Naturalis Principia Mathematica} in 1687\footnote{Robert Hooke famously claimed priority in the formulation of the inverse square law, but, as Alexis Clairaut wrote in 1759 concerning this dispute, there is a difference ``between a truth that is glimpsed and a truth that is demonstrated'' (quoted and translated in~\cite{Ball:1893fk}).}.  Thus, Kepler was neither the one who performed the original observations nor was he the one who discovered the explanation for the patterns exhibited by the observational data.  He was, in fact, performing precisely the same general function as statisticians, climate scientists, and anyone performing clinical drug trials: he was fitting the data to a mathematical structure; \emph{he was modeling}.  This is \emph{neither} theory \emph{nor} experiment.

To some extent we have, as scientists, successfully ignored this problem for four centuries largely because it didn't seem to matter.  After all, the dichotomy of theory and experiment was only a rough guide anyway and didn't have much of an impact (if any) on the science itself.  But now, in certain areas of science and particularly in physics, this dichotomy does not appear to be working as it should.  The most obvious example of this may be quantum mechanics where we have more than a century's worth of reliable experimental data, a well-established mathematical structure fit to that data, but no universally agreed upon \emph{interpretation} of this data and its mathematical structure.  Conversely, with string theory we have a well-established mathematical structure and a generally agreed-upon theory, but \emph{no data}.  In climate science, on the other hand, we have a consensus theory concerning climate change and we have a vast amount of experimental data, but we have no universally agreed upon mathematical model taking all of this data into account (i.e. we haven't reduced climate change to a self-contained set of equations yet).  These examples appear to suggest that Stodden is on the right track in suggesting that there is a third division to science.

But how would adding a third division of science to the usual two solve the problems raised by Lehrer, Andersen, and others?  To answer this question, let us first re-examine the purpose of each division's methods.  What is it that experimentalists are \emph{really} doing?  Are they actually describing the universe or is their aim something else?  I would argue that the aim of experimental science is, in fact, \emph{not} to merely describe the universe.  Even Aristotle described the universe.  What Aristotle \emph{didn't} do was describe it in a precise and consistent manner.  His interpretation of what he saw had to fit pre-conceived philosophical notions.  The revolution that marked the advent of modern \emph{experimental} science aimed at measuring quantities free from pre-conceived notions of what those quantities \emph{should} be.  In other words, experimental science does not describe things, it \emph{measures} things.  Inherent in this aim is \emph{precision} for measurement without precision is meaningless.  Achieving a measure of precision itself requires repeatability - experimental results must be repeatable and independently verifiable.  In fact, this latter point is so crucial that it is often \emph{more} important for experimentalists to describe their procedures as opposed to their data.  The data will often speak for itself but the procedure itself must be comprehensible if it is to be repeated and verified.

The aim of theory, on the other hand, has always been to \emph{explain} the world around us and not merely to describe it.  What sets \emph{modern} theoretical science apart from Aristotelianism and other historical approaches is that it aims for logical self-consistency with the crucial additional assumption that science, as a whole, is ultimately universal.  This last point implies that all of science is intimately \emph{connected}.  Thus we fully expect that biological systems, for example, will still obey physical and chemical laws.  Crucially, modern theoretical science also aims to \emph{predict} the future behavior of systems.  Thus a `good' scientific theory is both explanatory as well as predictive.

Description, then, is the realm of mathematics.  Mathematics is ultimately how we describe what we `see' in the experimental data.  However, since mathematics is \emph{such} an integral part of science, neither theorists nor experimentalists can carry out their work entirely free of it.  It is this all-pervasive nature of mathematics that then leads to confusions and mis-attributions of the kind argued by Lehrer as well as interpretational problems vis-\`{a}-vis probability theory and its relation to statistics.  As we noted earlier, roughly speaking, statistics generally is applied to prior knowledge (collected data) whereas probability theory is predictive in nature.  As such, statistics is generally descriptive whereas probability theory is predictively explanatory.  Thus I would argue that some of these issues could be cleared up if, rather than thinking of science in the way Alhazen did, perhaps with the added `third division' suggested by Stodden, we instead should think of science as being divided into three \emph{functions}: \textbf{measurement}, \textbf{description}, and \textbf{predictive explanation}.  These functions, of course, are the \emph{essence} of reductionism.

Now consider the rather sticky example of quantum mechanics which appears to be lacking a single, unifying `interpretation' (i.e. `theory' in the sense we have discussed above).  In our parlance, it would seem that there are multiple predictive explanations that exist for quantum mechanics.  But, in fact, most of the differences in the various interpretations of quantum mechanics differ in their interpretation of the quantum state.  Thus consider a generic quantum state,
$$
|\Psi\rangle = c_{1}|\psi_{1}\rangle + c_{2}|\psi_{2}\rangle.
$$
If we interpret this statistically, then the values $c_{1}$ and $c_{2}$ are arrived at only by making repeated measurements.  Instead, we can interpret this as a state of knowledge about the system that can be updated with a subsequent measurement.  In other words, it can be interpreted as being predictive, at least in a probabilistic sense.  On the other hand, if we take the state to be ontological, then it actually exists in the form given by $|\Psi\rangle$ and thus the state is merely descriptive.  Thus these three interpretations of the quantum state correspond exactly to the three `functions' of science and, when viewed in that light, do not necessarily contradict one another.  Perhaps, instead of requiring \emph{no} interpretation, as Brukner has suggested~\cite{Swarup:2009fk}, quantum mechanics actually requires \emph{multiple} interpretations.

Finally, if we then return to the problems of complexity and emergence, if science is to be considered universal, connective, and self-consistent, perhaps the problem is not that reductionism is a broken paradigm, but rather that we are mis-ascribing some of our activities to the wrong scientific function, e.g. perhaps some of our so-called theories are actually more descriptive than predictively explanatory.  Or perhaps they're built on the wrong description.  Either way, it may be a bit premature to declare reductionism dead.  In fact it may simply be that, since the time of Alhazen, we have simply been missing a key ingredient.  In order to maintain science as a productive, respected, and vital discipline we must ensure that it remains true to its foundational functions and always allows for introspection.  Otherwise, science risks being ignored and too much is at stake for us to let that happen.

\begin{acknowledgments}
The ideas discussed in this essay were tested on a few unsuspecting audiences over the course of a little more than a month.  Thus, for helping me feel my way through these ideas, I would like to thank the following for giving me a pulpit from which to preach: the Clemson University Symposium for Introduction to Research in Physics and Astronomy (SIRPA); the Kennebunk Free Library's `Astronomy Nights,' co-hosted by the Astronomical Society of Northern New England; and the Saint Anselm College Philosophy Club.
\end{acknowledgments} 

\newpage

\bibliographystyle{plain}
\bibliography{FQXi3.bib}
\end{document}